\documentclass[aps,pre,twocolumn,superscriptaddress,longbibliography]{revtex4-2}

\usepackage{amssymb}
\usepackage{amsmath}
\usepackage{bm}
\usepackage{color}
\usepackage{graphicx}
\usepackage{hyperref}
\usepackage{braket}
\usepackage{placeins}

\newcommand{\lla}{\left\langle}
\newcommand{\rra}{\right\rangle}

\DeclareFontFamily{U}{mathc}{}
\DeclareFontShape{U}{mathc}{m}{it}
{<->s*[1.03] mathc10}{}

\DeclareMathAlphabet{\mathscr}{U}{mathc}{m}{it}

\DeclareMathOperator*{\equals}{=}

\graphicspath{{./figures/}, {./supplement/figures}}

\begin{document}

\title{Dynamics of Active Polar Ring Polymers}

\author{Christian A. Philipps}
\email{c.philipps@fz-juelich.de}
\affiliation{Theoretical Physics of Living Matter, Institute of Biological Information Processing and Institute for Advanced Simulation, Forschungszentrum J\"ulich and JARA,
	52425 J\"ulich, Germany}
	\affiliation{Department of Physics, RWTH Aachen University, 52056 Aachen, Germany}
\author{Gerhard Gompper}
\email{g.gompper@fz-juelich.de}
\affiliation{Theoretical Physics of Living Matter, Institute of Biological Information Processing and Institute for Advanced Simulation, Forschungszentrum J\"ulich and JARA,
	52425 J\"ulich, Germany}
\author{Roland G. Winkler}
\email{r.winkler@fz-juelich.de}
\affiliation{Theoretical Physics of Living Matter, Institute of Biological Information Processing and Institute for Advanced Simulation, Forschungszentrum J\"ulich and JARA,
	52425 J\"ulich, Germany}

\begin{abstract}

The conformational and dynamical properties of isolated semiflexible active polar ring polymers are investigated analytically. A ring is modeled as continuous Gaussian polymer exposed to tangential active forces. The analytical solution of the linear non-Hermitian equation of motion in terms of an eigenfunction expansion shows that ring conformations are independent of activity. In contrast, activity strongly affects the internal ring dynamics and yields characteristic time regimes, which are absent in passive rings. On intermediate time scales, flexible rings show an activity-enhanced diffusive regime, while semiflexible rings exhibit ballistic motion. Moreover, a second active time regime emerges on longer time scales,  where rings display a snake-like motion, which  is reminiscent to a tank-treading rotational dynamics in shear flow, dominated by the mode with the longest relaxation time.

\end{abstract}

\maketitle

\section{Introduction} 

Filaments and polymers are  fundamental ingredients of living matter and essential for the diverse functions of eukaryotic and prokaryotic cells. The out-of-equilibrium processes in these cells affect the conformations and dynamics of the immanent polymeric structures.  Molecular motors walking along microtubule filaments generate forces that determine the dynamics of the cytoskeletal network and the organization of the cell interior \cite{lau_microrheology_2003,mackintosh_nonequilibrium_2008,lu_microtubulemicrotubule_2016,ravichandran_enhanced_2017}. Even more, molecular motors give rise to nonequilibrium conformational fluctuations of actin filaments and microtubules \cite{brangwynne_cytoplasmic_2008,weber_random_2015}.
Within the nucleus, ATPases such as DNA or RNA polymerase (DNAP and RNAP, respectively) are involved in DNA transcription and every RNAP/DNAP translocation step  generates nonthermal fluctuations for both RNAP/DNAP and the transcribed DNA \cite{guthold_direct_1999,mejia_trigger_2015,belitsky_stationary_2019,winkler_physics_2020}. Among the wide spectrum of polymeric structures,  chromosomes in bacteria \cite{wu_direct_2019}, archaea, chloroplasts, and even mitochondrial \cite{koche_extrachromosomal_2020} and extrachromosomal DNA \cite{cao_extrachromosomal_2021} of eukaryote cells are of circular nature, similarly, actomyosin aggregates in cytokinesis and marginal bands formed by microtubules in blood cells   \cite{sehring_assembly_2015,cheffings_actomyosin_2016,dmitrieff_balance_2017}. The circular shape substantially affects their dynamical behavior which deviates from that of comparable linear structures.  This is emphasized in experiments on microtubules placed on motility assays, which reveal ring-like structures \cite{kawamura_ring-shaped_2008,liu_loop_2011,keya_synchronous_2020} and an emergent rotational motion \cite{kawamura_ring-shaped_2008,keya_synchronous_2020}.

The desire to unravel the underlying physical phenomena and to gain insight into the emergent behaviors of the out-of-equilibrium  polymeric structures has prompted intensive studies on tangentially (active polar)  \cite{ravichandran_enhanced_2017,winkler_physics_2020,peruani_nonequilibrium_2006,abkenar_collective_2013,bar_self-propelled_2020,liverpool_viscoelasticity_2001,isele-holder_self-propelled_2015,bianco_globulelike_2018,anand_structure_2018,locatelli_activity-induced_2021,peterson_statistical_2020,winkler_physics_2020} and isotropically (active Brownian) driven or self-propelled linear filaments and polymers \cite{ghosh_dynamics_2014,harder_activity-induced_2014,kaiser_unusual_2014,kaiser_how_2015,chaki_enhanced_2019,shin_facilitation_2015,eisenstecken_conformational_2016,eisenstecken_internal_2017,martin-gomez_active_2019,anand_conformation_2020,martin-gomez_hydrodynamics_2020,mousavi_active_2019,winkler_physics_2020} --- so-called active polymers. Even experiments on living worms as a model system resembling tangentially driven polymers have been performed \cite{deblais_phase_2020}. These studies reveal a  strong influence of the active forces on the polymer conformations and dynamics, and can lead to polymer swelling \cite{ghosh_dynamics_2014,kaiser_unusual_2014,eisenstecken_conformational_2016} or shrinkage \cite{isele-holder_self-propelled_2015,bianco_globulelike_2018,anand_structure_2018,anand_conformation_2020,locatelli_activity-induced_2021}, depending on the kind of active force and the environment \cite{isele-holder_self-propelled_2015,bianco_globulelike_2018,martin-gomez_active_2019,martin-gomez_hydrodynamics_2020,winkler_physics_2020}. 

Active polymers are typically studied by computer simulations employing various discrete models \cite{winkler_physics_2020}, which differ in the way the tangential forces are applied, yet yield the same continuum limit for smooth contours \cite{bianco_globulelike_2018,anand_structure_2018,locatelli_activity-induced_2021}. However, a severe problem arises for flexible discrete polymers, where the bending angles are not restricted and can become very large, which renders the definition of a tangent vector arbitrary --- related to the well-known property of random walks to generate  non-differentiable trajectories. Moreover, analytical descriptions and results, which could serve as a guide to uncover model-specific discretization phenomena, for 
active polar ring polymers are lacking \cite{peterson_statistical_2020}.

\begin{figure}[t]
	\centering
	\includegraphics[width=0.9\columnwidth]{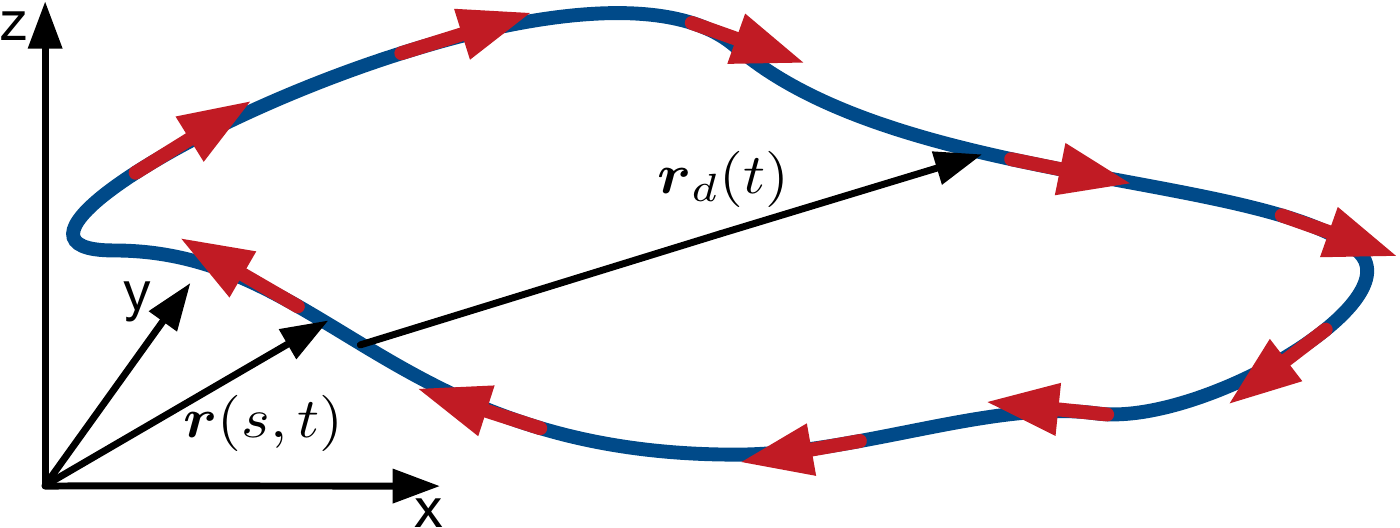}
	\caption{Illustration of an active polar ring polymer. The arrows indicate the local tangential active force. An arbitrary position vector $\bm{r}(s,t)$ and the ring diameter vector $\bm{r}_d(t)$ are shown.
	}
	\label{fig::Sketch_RPM} 
\end{figure}

To address this fundamental question, we present analytical results for the conformations and dynamics of continuous phantom semiflexible active polar ring polymers (APRPs), based on a Gaussian polymer model \cite{winkler_deformation_2003,harnau_dynamic_1995}. Due to the activity, the equation of motion is non-Hermitian, and the solution in terms of an eigenfunction representation yields complex eigenvalues.  The latter imply a particular internal ring dynamics, distinctly different from that of passive rings, with a snake-like motion along their contour for all stiffnesses,  which is similar to a tank-treading rotational motion known for passive ring polymers \cite{chen:13.2,lieb:18.1} and vesicles \cite{kell:82,nogu:04} under shear flow. Hence,  we denote this  dynamical behavior as {\em active tank-treading}. This is reflected in characteristic time regimes, with an enhanced diffusive and a ballistic dynamics for flexible and semiflexible rings, respectively.
We find marked qualitative and quantitative differences to the results of previous simulations of a discrete model \cite{locatelli_activity-induced_2021}, like an average ring size, which is found to be independent of activity in our model, but has been predicted to display a very strong swelling in Ref.~\cite{locatelli_activity-induced_2021}. We also present results of a modified discrete model \cite{isele-holder_self-propelled_2015}, which yields qualitatively agreement with our continuum predictions.

\section{Model}

The ring polymer is described as a continuous, differentiable space curve $\bm{r}(s,t)$ embedded in three dimensions, where $s \in [0,L)$ is the contour variable along the ring and $t$ the time (Fig.~\ref{fig::Sketch_RPM}). We adopt the Gaussian semiflexible polymer model \cite{winkler_deformation_2003,harnau_dynamic_1995}, which yields the overdamped --- the inertia term is omitted --- equation of motion  for an APRP 
\begin{align} 
	\gamma  \frac{\partial \bm{r}(s,t)}{\partial t} = \: &2k_BT \lambda \frac{\partial^2 \bm{r}(s,t)}{\partial s^2} - k_BT  \epsilon \frac{\partial^4\bm{r}(s,t)}{\partial s^4} + \bm{\Gamma}(s,t)
	\nonumber
	\\
	& + f_a \frac{\partial \bm{r}(s,t)}{\partial s},
	\label{eq::EOM}
\end{align}		
where $\gamma$ is the translational friction coefficient per length, $k_B$ the Boltzmann constant, and $T$ the temperature (cf. Supplemental Material \cite{philipps_supplemental_2022}). 
The bending rigidity is given by $\epsilon = 3/(4p)$, with $p=1/(2l_p)$, in terms of the persistence length $l_p$. The Lagrangian multiplier $\lambda$ is determined by the inextensibility of the ring contour via the local constraint of the mean-square tangent vector  $\langle (\partial \bm{r}(s,t) / \partial s)^2 \rangle = 1$ \cite{eisenstecken_conformational_2016, mousavi_active_2019}.  The homogeneous external or internal  active force of magnitude $f_a$ acts in the direction of the local tangent $\partial \bm{r}(s,t)/ \partial s$.  
Thermal fluctuations are captured by the stochastic force $\bm{\Gamma}(s,t)$, which  is assumed to be stationary, Markovian, and Gaussian with zero mean and second moment $\langle \bm \Gamma (s,t) \cdot \bm \Gamma (s',t') \rangle = 6 \gamma k_BT \delta(s-s') \delta(t-t')$. 

The linear, but non-Hermitian, Langevin equation \eqref{eq::EOM} is solved by the eigenfunction expansion $\bm{r}(s,t) =\sum_{m=-\infty}^{\infty} \bm{\chi}_m(t) \phi_m(s)$ with the eigenfunctions $ \phi_m(s) = e^{i k_m s} / \sqrt{L}$ and the wave numbers $k_m = 2\pi m/L$, which satisfy the periodic boundary condition $\bm{r}(s,t) = \bm{r}(s+L,t)$ by the ring structure. Insertion of the expansion into Eq.~\eqref{eq::EOM} yields the equations of motion for the normal mode amplitudes, $\bm \chi_m(t)$, 
\begin{equation}
	\gamma \frac{d}{d t} \bm{\chi}_m(t) = - \xi_m \bm{\chi}_m(t) + \bm{\Gamma}_m(t) .
	\label{eq::temporal_EOM}
\end{equation}	
The eigenvalues $\xi_m$ of the eigenvalue problem are given by ($m \in \mathbb{Z}$) 
\begin{equation}
	\xi_m = \frac{12\pi^2k_BTpL}{L^3} \left[ \frac{\pi^2}{(pL)^2}m^4 + \mu m^2 - i \frac{Pe}{6\pi pL} m \right] ,
\end{equation}	
with the abbreviation $\mu = 2 \lambda/(3p)$ and the P\'eclet number 
\begin{equation}	
	Pe = \frac{f_aL^2}{k_BT} ,
	\label{eq::pe}
\end{equation}
which characterizes the strength of the activity. The $\xi_m$ are complex due to the non-Hermitian nature of the underlying equation with the first-order derivative, which implies a dynamical behavior absent in passive systems and active polar linear polymers. The correlation functions of the normal-mode amplitudes determine the ring dynamics and capture the influence of activity. Explicitly, they are given in the stationary state by ($m \neq 0$)
\begin{equation} 
    \hspace{-0.1cm}
	\lla \bm{\chi}_m(t) \cdot \bm{\chi}_n^{*}(t') \rra = \delta_{mn} \frac{3 k_BT}{\gamma} \tau_m e^{-|t-t'|/\tau_m} e^{i \omega_m|t-t'|} ,
	\label{eq::AMCF}
\end{equation}	
where $\bm{\chi}_m^{*}(t) = \bm{\chi}_{-m}(t)$. Here, the eigenvalues $\xi_m = \xi_m^R - i \xi_m^I$ are separated in  real and imaginary parts, and the relaxation times $\tau_m = \gamma / \xi_m^{R}$ and frequencies $\omega_m = \xi_m^{I} / \gamma = 2\pi f_a m/(\gamma L)$ are introduced, where $\tau_m$ is activity independent and identical with the relaxation time of a passive ring \cite{mousavi_active_2019}. Most importantly, the non-Hermitian nature of a ring's equation of motion implies complex correlation functions with an activity-dependent frequency. Noteworthy, at equal times $t=t'$, the active contribution in Eq.~\eqref{eq::AMCF} vanishes and $\langle \bm{\chi}_m(t) \cdot \bm{\chi}_n^{*}(t) \rangle$ is equal to the equilibrium  correlation function of a  passive ring. This illustrates that tangential active propulsion of a continuous polymer only impacts the ring dynamics, but not its conformations.  

\section{Dynamics}

The translational motion of a ring is characterized by its mean-square displacement (MSD), $\langle \Delta \bm r^2_{tot}(t) \rangle = \langle (\bm r(s,t) - \bm r(s,0) )^2 \rangle$, which can be separated into a contribution from the center-of-mass motion, $\langle \Delta \bm r^2_{cm}(t) \rangle$, and a ring internal part, $\langle \Delta \bm r^2(t) \rangle$, such that $
	\langle \Delta \bm r^2_{tot} (t) \rangle = \langle \Delta \bm r^2_{cm}(t) \rangle + \langle \Delta \bm r^2(t) \rangle$.
Due to the ring structure, the MSD is independent of the contour variable $s$. Moreover, by integrating the Langevin equation \eqref{eq::EOM} over the ring contour, all internal and active forces vanish, and the center-of-mass diffusion is solely determined by thermal fluctuations with the diffusion coefficient  $D_0=k_BT/(\gamma L)$ of a passive ring. This is at variance to the model applied in Ref.~\cite{locatelli_activity-induced_2021}, where the sum of active forces along the ring polymer is non-zero. 
On the contrary, in tangentially driven active polar linear polymers, the sum over the active forces must obviously be large, and an activity-enhanced long-time diffusion coefficient is obtained \cite{isele-holder_self-propelled_2015,bianco_globulelike_2018, peterson_statistical_2020}.  

The MSD in the center-of-mass reference frame is given by 
\begin{equation} 
	\lla \Delta \bm r^2(t) \rra = \frac{12 k_BT}{\gamma L} \sum_{m=1}^{\infty} \tau_m \left(1- \cos(\omega_m t)  e^{-t/\tau_m} \right).
	\label{eq::msd_inter}
\end{equation}
Compared to a passive ring, with $Pe=0$ \eqref{eq::pe}, an additional periodic function, $\cos(\omega_m t)$, appears for active rings, which determines the MSD over certain time scales $t/\tau_1 \lesssim 1$, where $\tau_1$ is the longest relaxation time. Equation~\eqref{eq::msd_inter} reveals two relevant time scales, the longest polymer relaxation time $\tau_1$ and the oscillation period by the lowest frequency $\omega_1$.  With the relaxation time  $\tau_1= \gamma L^3/(12 \pi^2 k_B T pL)$ of flexible and $\tau_1= \gamma L^3/(24 \pi^2 k_B T)$ of semiflexible rings \cite{mousavi_active_2019}, the relation
\begin{align}
	\omega_1 \tau_1 \approx \left\{ 
	\begin{array}{cc} 
		\displaystyle \frac{\displaystyle Pe}{\displaystyle 6 \pi pL}  \ , &  pL \gg 1  \\[10pt]
		\displaystyle \frac{\displaystyle Pe}{\displaystyle 12 \pi}  \ , &   pL \ll 1
	\end{array} 
	\right.
	\label{eq::omega_relax} 
\end{align} 
is obtained, which determines the importance of the cosine term.

\begin{figure}[t]
	\centering
	\includegraphics[width=0.95\columnwidth]{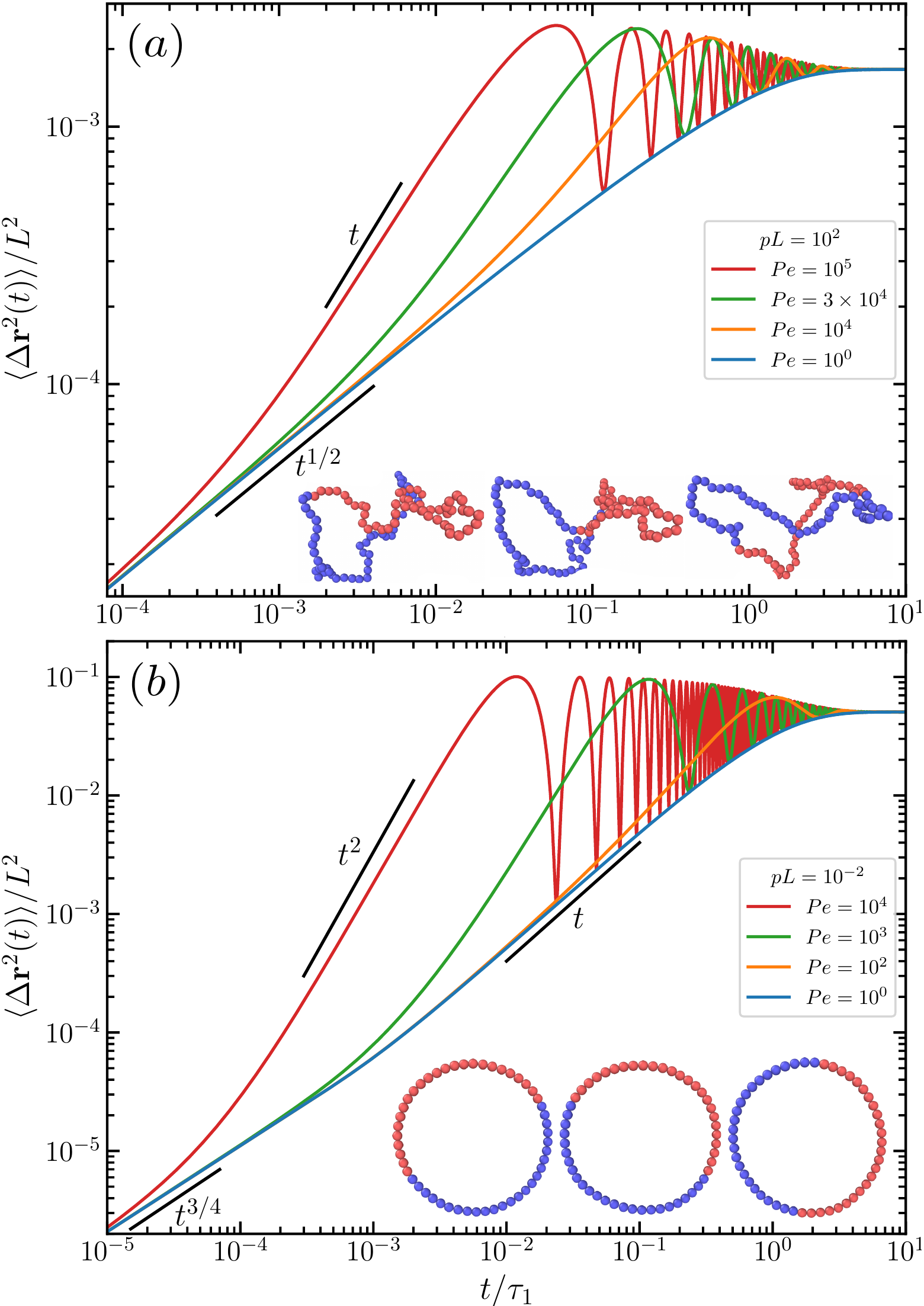}
	\caption{Normalized mean-square displacement in the center-of-mass reference frame, $\langle \Delta \bm{r}^2(t) \rangle$, of $(a)$ flexible, $pL=10^2$, and $(b)$ semiflexible, $pL=10^{-2}$, APRPs as function of the time $t/\tau_1$, where $\tau_1$ is the longest relaxation time, for various Péclet numbers $Pe$. The black lines show power laws with the indicated time dependence. The insets show subsequent conformations with $\Delta t \approx 1/ \omega_1$ of discrete flexible and semiflexible polymers of length $L=50l$. To illustrate the clockwise  tank-treading motion, half of the monomers are colored blue and red, respectively.  }
	\label{fig::RPM_con_act_flex_semi_MSD}
\end{figure}	

Figure~\ref{fig::RPM_con_act_flex_semi_MSD} displays MSDs of flexible and semiflexible APRPs for various P\'eclet numbers.
As shown in Fig.~\ref{fig::RPM_con_act_flex_semi_MSD}(a), on time scales $t/\tau_1 \ll 1/(\omega_1 \tau_1)^2 \ll 1$, flexible polymers exhibit the well-known sub-diffusive time dependence $t^{1/2}$ predicted by the Rouse model \cite{doi_theory_1986,hur_comparison_2006}. In the range $1/(\omega_1 \tau_1)^2 < t/\tau_1 \ll 1/(\omega_1 \tau_1)$ an activity-enhanced linear time regime appears for  $Pe \gtrsim 10^4$, where 
\begin{equation} 
	\langle \Delta \bm{r}^2(t) \rangle \approx \frac{L^2}{12\pi^2} \frac{Pe}{(pL)^2} \frac{t}{\tau_1} ,
	\label{eq::msd_approx_flex}
\end{equation}
with a linear $Pe$ dependence.
For longer times, $1/(\omega_1 \tau_1)< t/\tau_1 \lesssim 1$, oscillations due to the cosine term emerge.
Here, all modes contribute to the MSD \cite{philipps_supplemental_2022}.  

Figure~\ref{fig::RPM_con_act_flex_semi_MSD}(b) for semiflexible polymers reveals a qualitatively similar behavior, with the characteristic  time dependence $t^{3/4}$ \cite{farge_dynamic_1993,winkler_diffusion_2007} for $t/\tau_1 \ll \mathrm{min}\{pL, (pL)^{1/5}/(\omega_1 \tau_1)^{8/5}\}$. In the range of $(pL)^{1/5}/(\omega_1 \tau_1)^{8/5}< t/\tau_1 \ll 1/(\omega_1 \tau_1)$,  the MSD is dominated by the first mode, which yields  $\langle \Delta \bm r^2 (t) \rangle \sim t$ for $Pe \ll 1$ \cite{winkler_diffusion_2007,philipps_supplemental_2022}, whereas for large P\'eclet numbers the  active ballistic time regime   
\begin{equation} 	
	\lla \Delta \bm{r}^2(t) \rra  \approx \frac{L^2 Pe^2}{576\pi^4}  \left(\frac{t}{\tau_1} \right)^2 
	\label{eq::msd_approx_semiflex}
\end{equation}	
emerges \cite{philipps_supplemental_2022}.
Here, the MSD shows a quadratic dependence on the P\'eclet number and is independent of persistence length. 
At  times $t/\tau_1  \gtrsim 1/(\omega_1 \tau_1)$, $\langle \Delta \bm r^2(t) \rangle$ is well described by 
\begin{equation}
	\lla \Delta \bm r^2(t) \rra \approx \frac{L^2}{2\pi^2} \left[ 1-\cos(\omega_1 t) e^{-t/\tau_1} \right] ,
\end{equation}
and oscillations appear.  

The difference in the $Pe$ dependence between flexible and semiflexible active polar ring polymers reflects the underlying distinctive conformations. Flexible polymers are coiled and all modes contribute to the internal dynamics. In contrast, semiflexible APRPs assume circular 
conformations and their dynamics on time scales $t/\tau_1 > 1/(\omega_1 \tau_1)$ is described by the mode with the longest relaxation time corresponding to a rotational motion. 

The oscillations appear as long as $\omega_1 \tau_1 >1$, i.e., the polymer relaxation time is longer than the period by the frequency $\omega_1$. This is reflected in Fig.~\ref{fig::RPM_con_act_flex_semi_MSD}, which illustrates that the  oscillations disappear with decreasing  $Pe$. Notice that the polymer relaxation time is independent of $Pe$, but $\omega_1$ is, which stresses the active nature of the effect.  

\begin{figure}[t!]
	\centering
	\includegraphics[width=0.95\columnwidth]{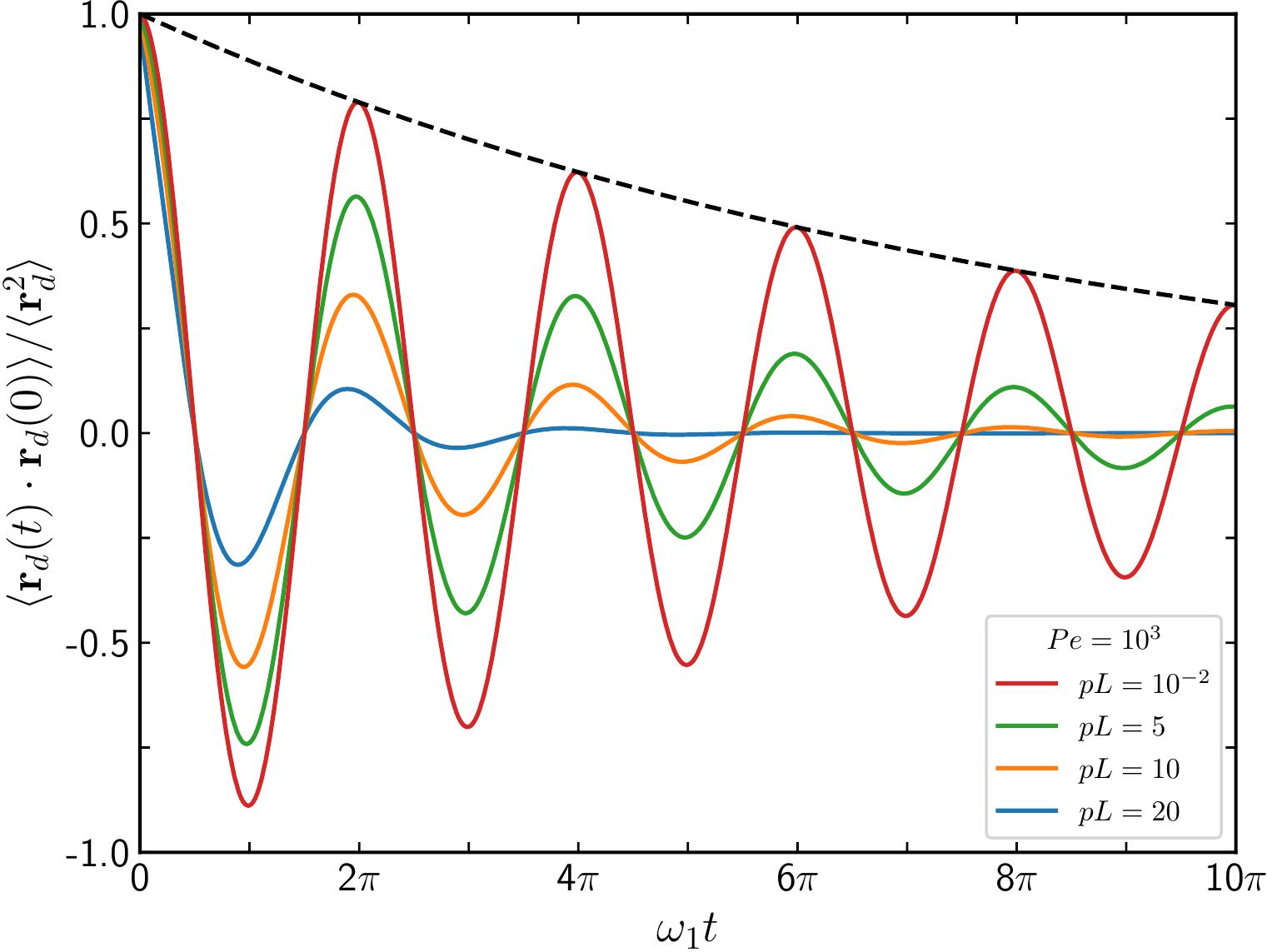}
	\caption{Normalized autocorrelation function of the ring diameter $\bm{r}_d(t)$ as a function of the time $\omega_1 t$, where $\omega_1 = 2 \pi f_a/(\gamma L)$, for $Pe =10^3$ and various $pL$ values as indicated in the legend. The black dashed line corresponds to the exponential $\exp(-t/\tau_1) = \exp(-12 \pi \omega_1 t/Pe)$. Curves for $pL <10^{-2}$ are indistinguishable from that with $pL =10^{-2}$.}
	\label{fig::correlation}
\end{figure}

To characterize the oscillations in the ring polymer dynamics, the temporal autocorrelation function of the ring diameter vector $\bm{r}_d(t) = \bm{r}(L/2,t)-\bm{r}(0,t)$ is considered (Fig. \ref{fig::Sketch_RPM}). Analytically, its correlation function is given by
\begin{equation}
	\langle \bm{r}_d(t) \cdot \bm{r}_{d}(0) \rangle = \frac{24k_BT}{\gamma L} \sum_{\substack{m=1 \\ m \: odd}}^{\infty} \tau_{m} \: \cos( \omega_{m} t) \: e^{-t / \tau_{m}} .
	\label{eq::corr}
\end{equation}	
The cosine term yields significant contributions as long as $\omega_1 \tau_1 >1$ on time scales $t/\tau_1 \lesssim 1$. As shown in Eq.~\eqref{eq::omega_relax},  
the product $\omega_1 \tau_1$ increases linearly with increasing $Pe$ for any stiffness, hence, oscillations always appear for sufficiently large P\'eclet numbers. Noteworthy, $\omega_1 \tau_1$ for $pL \ll 1$ is independent of ring stiffness and length. This yields a universal dynamical behavior in terms of $\omega_1 \tau_1$ for $t/\tau_1 <1$. For flexible polymers, $\omega_1 \tau_1$ depends on $pL$, and the P\'eclet number has to exceed the value of $6 \pi pL$ to observe oscillations, which requires much larger P\'eclet numbers compared to semiflexible rings. As shown in Figure~\ref{fig::correlation}, for $Pe=10^3$ the correlation function already decays exponentially for $pL \gtrsim 50$. 

In the regime of pronounced oscillations, the correlation function for all ring stiffnesses is determined by the first mode in Eq.~\eqref{eq::corr}, i.e.,
\begin{equation}
	\frac{\lla \bm{r}_d(t) \cdot \bm{r}_{d}(0) \rra}{\lla \bm{r}_d^2 \rra} \approx C_\omega \cos\left( \omega_1 t \right) e^{-t/\tau_1}, 
\end{equation} 
with the equilibrium mean-square ring diameter $\langle \bm r_d^2 \rangle$ \cite{mousavi_active_2019,philipps_supplemental_2022} of the passive ring and a stiffness-dependent constant $C_\omega$. In case of a flexible polymer, the amplitude is $C_\omega =8/\pi^2$, and its deviation from unity reflects the influence of higher modes. In fact, the deviation is below $20\%$ for all $pL$ values.  The ring performs an active tank-treading motion along the slowly varying instantaneous conformation when $Pe > 6 \pi pL$, with the frequency $\omega_1$, corresponding to the tangential velocity  $f_a/\gamma$ (cf. Supplemental Material,  movie M1 and movie M2 \cite{philipps_supplemental_2022}). 

The correlation function of semiflexible rings, $pL < 1$, is governed by the first mode only and the longest relaxation time is independent of $pL$, hence,  $C_\omega =1$.  The damped periodic dynamics corresponding to  a tank-treading rotation of the ring with the frequency $\omega_1$ (cf. Supplemental Material,  movie M3 \cite{philipps_supplemental_2022}). In general, activity implies a tank-treading motion and the thermal (passive) contributions control the damping. 

\section{Discussion}

Our analytical studies of continuous semiflexible active polar ring polymers reveal substantial differences compared to simulations of rings modeled by discrete points, which is related to the definition of the active force.  In Ref.~\cite{locatelli_activity-induced_2021}, the tangential force $\tilde{\bm F}_i = F_a \bm t_i$, with the unit vector $\bm t_i = (\bm r_{i+1} - \bm r_{i-1})/|\bm r_{i+1} - \bm r_{i-1}|$ and strength $F_a$, on particle $i$ is applied, where $\bm r_{i \pm 1}$ are the positions of the neighboring particles along the ring contour. For any discrete ring, this active force differs from the alternative discretization $ \bm F_i = F_a (\bm r_{i+1}  - \bm r_{i-1})/(2l)$, with the bond length $|\bm r_{i+1}  - \bm r_{i}| =l$, which is the sum of the forces $\bm F_i = F_a (\bm R_{i+1}+\bm R_i)/(2l)$ along the two bond vectors, where $\bm R_i = \bm r_i - \bm r_{i-1}$ (cf. Supplemenatray material \cite{philipps_supplemental_2022}) \cite{isele-holder_self-propelled_2015}. As the angle between two subsequent bonds changes, $\bm F_i$ varies strongly --- it assumes a maximum for parallel and vanishes for antiparallel bond alignment. In contrast, the force $\tilde{\bm F}_i$ is independent of the angle between the successive bonds. In both cases, the force is ``tangential'' to the contour, and turns into the active force in Eq.~\eqref{eq::EOM} in the continuum limit. 

Most importantly, the conformations of the continuous rings are independent of activity. This is in stark contrast to simulations based on the tangential force $\tilde{\bm F}_i$ \cite{locatelli_activity-induced_2021}, which predict a strong swelling of phantom polymers with increasing activity by approximately $250\%$ of the mean radius-of-gyration at the P\'eclet number $Pe=F_a l/(k_BT) =10$ (cf. Supplemental Material of Ref.~\cite{locatelli_activity-induced_2021}). We have performed Brownian dynamics simulations of flexible and semiflexible APRPs applying the active forces $\bm F_i$ above to resolve the fundamental difference in the ring conformations (for details see Ref.~\cite{philipps_supplemental_2022}). These simulations for flexible rings, with $Pe=20$, yield a small shrinkage of the ring by approximately $10\%$ of the root mean-square radius-of-gyration with respect to its value at equilibrium. This constitutes a very minor conformational change compared to that observed in Ref.~\cite{locatelli_activity-induced_2021}, and is consistent with our analytical results. The discrepancy reveals a strong influence of the discretization of the active force on the ring conformations. In addition, the integral over the active force in Eq.~\eqref{eq::EOM} is zero, i.e., a ring's center-of-mass dynamics is independent of tangential propulsion. Again, this is in contrast to simulations applying the active force of Ref.~\cite{locatelli_activity-induced_2021}.  

Experimental realizations of synthetic active rings using self-propelled Janus particles show variations in the Janus-particle orientations, and their propulsion directions are not always necessarily tangential \cite{nishiguchi_flagellar_2018}. However, biological filaments such as microtubules, actin filaments, and circular chromosomes need to be described by a semiflexible polymer model, at least on a local scale, and the difference between the various discretization schemes is expected to be of minor importance. We expect our theoretical approach to capture the essential features of such active polar ring polymers. 

The discussed dynamical properties of APRPs should be experimentally accessible via structures formed by microtubules \cite{dmitrieff_balance_2017,liu_loop_2011,keya_synchronous_2020} or actin filaments driven by molecular motors. For a microtubule of length $L=1 \, \mathrm{\mu m}$, the force per motor $F_a =  6 \, \mathrm{pN}$, and $N = 10$ active motors,  the P\'eclet number at room temperature is $Pe = N\, F\, L/(k_BT) \approx  10^4$  \cite{rupp_patterns_2012}, on the order of the P\'eclet numbers in Figures~\ref{fig::RPM_con_act_flex_semi_MSD} and \ref{fig::correlation}. Experiments on microtubules placed on motility assays indeed exhibit  rotational motion \cite{kawamura_ring-shaped_2008,keya_synchronous_2020}, consistent with our prediction. Tank-treading motion can also be expected for circular aggregates of crosslinked microtubules \cite{dmitrieff_balance_2017} or actin filaments. Such structures can be synthesized and would provide, in combination with motility assays, insight into the nonequilibrium dynamical properties of flexible and semiflexible APRPs.       

We have focused on the dynamical properties of idealized active rings. Passive rings in a melt exhibit strong conformational changes and shrinkage with increasing concentration \cite{kapnistos_ring_melt_2008,reigh_ring_2013}. Here, self-avoidance and excluded-volume interactions with  and entanglements by the surrounding ring polymers play a major role. It is not a priori evident, how the conformations of active rings are affected in this case. However, the predicted  active tank-treading motion will certainly be present.

\section*{Acknowledgements}

We would like to thank  J. Midya and G. A. Vliegenthart for constructive discussions.


%

\newpage 

\begin{widetext}

\setcounter{equation}{0}
\setcounter{section}{0}
\setcounter{figure}{0}

\renewcommand{\theequation}{S\arabic{equation}}
\renewcommand{\thefigure}{S\arabic{figure}}
\renewcommand{\thesection}{S\arabic{section}}

\begin{center} {\large {\bf Supplemental Material \\[5pt]
 Dynamics of Active Polar Ring Polymers}}\\[2ex]  
 Christian A. Philipps \\
 {\em Theoretical Physics of Living Matter, Institute of Biological Information Processing and Institute for Advanced Simulation, Forschungszentrum J\"ulich and JARA,
  	52425 J\"ulich, Germany  and  \\ Department of Physics, RWTH Aachen University, 52056 Aachen, Germany} \\[1ex]
  Gerhard Gompper and Roland G. Winkler \\
   {\em Theoretical Physics of Living Matter, Institute of Biological Information Processing and Institute for Advanced Simulation, Forschungszentrum J\"ulich and JARA,
    	52425 J\"ulich, Germany}
\end{center}

\author{Christian A. Philipps}
\email{c.philipps@fz-juelich.de}
\affiliation{Theoretical Physics of Living Matter, Institute of Biological Information Processing and Institute for Advanced Simulation, Forschungszentrum J\"ulich and JARA,
	52425 J\"ulich, Germany}
	\affiliation{Department of Physics, RWTH Aachen University, 52056 Aachen, Germany}
\author{Gerhard Gompper}
\email{g.gompper@fz-juelich.de}
\affiliation{Theoretical Physics of Living Matter, Institute of Biological Information Processing and Institute for Advanced Simulation, Forschungszentrum J\"ulich and JARA,
	52425 J\"ulich, Germany}
\author{Roland G. Winkler}
\email{r.winkler@fz-juelich.de}
\affiliation{Theoretical Physics of Living Matter, Institute of Biological Information Processing and Institute for Advanced Simulation, Forschungszentrum J\"ulich and JARA,
	52425 J\"ulich, Germany}

\maketitle

\section{Equation of Motion}

The dynamics of the ring polymer is described by the overdamped Langevin equation of motion (EOM)
\begin{equation} 
	\gamma  \frac{\partial \bm{r}(s,t)}{\partial t} = f_a \frac{\partial \bm{r}(s,t)}{\partial s} + \: 2k_BT \lambda \frac{\partial^2 \bm{r}(s,t)}{\partial s^2} - \epsilon k_BT \frac{\partial^4\bm{r}(s,t)}{\partial s^4} + \bm{\Gamma}(s,t) ,
	\label{eq::EOM}
\end{equation}
accounting for the conformational degrees of freedom, second term on the right-hand side, and bending restrictions, third term on the right-hand side,  where $\epsilon = 3/(4p)$, $p=1/(2l_p)$, and $l_p$ is the persistence length. The Lagrangian multiplier $\lambda$ respects the inextensibility of the ring, and $k_BT$ is the thermal energy [S1,S2]. $\bm \Gamma (s,t)$ captures thermal fluctuations, and $f_a$ is the strength of active force per unit length along the local tangent of the ring. The EOM \eqref{eq::EOM} is solved by an eigenfunction expansion,
\begin{equation}
    \bm{r}(s,t) = \sum_{m=-\infty}^{\infty} \bm{\chi}_m(t) \phi_m(s) ,
    \label{eq::eigenfunction_expansion}
\end{equation}
which leads to the eigenvalue problem
\begin{equation}
	\sigma \phi_m(s) = - \xi_m \phi_m(s) .
	\label{eq::spatial_EOM}
\end{equation}	
The non-Hermetian differential operator $\sigma$ is given by $\sigma = f_a \partial/\partial s + 2k_BT\lambda \partial^2/\partial s^2 - \epsilon k_BT \partial^4/\partial s^4 \ne \sigma^{\dagger}$. The complex eigenvalues $\xi_m = \xi_m^{R} - i \xi_m^{I}$ and the complete set of orthogonal eigenfunctions $\phi_m(s)$ are
\begin{align}
    \xi_m &= \frac{12\pi^2k_BTpL}{L^3} \left[ \frac{\pi^2}{(pL)^2}m^4 + \mu m^2 - i \frac{Pe}{6\pi pL} m \right] ,
    \\
    \phi_m(s) &= \frac{1}{\sqrt{L}} e^{ik_ms} ,
\end{align}
with the wave numbers $k_m = 2 \pi m / L$, $m \in \mathbb{Z}_{0}$, the relaxation times $\tau_m$, and the frequencies $\omega_m$: 
\begin{align}
    \tau_m &= \frac{\gamma}{\xi_m^R} = \frac{\gamma L^3}{12 \pi^2 k_B T pL \left(\displaystyle \pi^2 m^4 /(pL)^2 + \mu m^2  \right)} \ , 
    \\  
    \omega_m  &= \frac{\xi_m^{I}}{\gamma} = \frac{2 \pi f_a m}{\gamma L} .  
\end{align}
Figure \ref{fig::RPM_con_pas_semi_tau} displays relaxation times for various $pL$ values. 

The solutions of the equations of motion for the normal-mode amplitudes, 
\begin{equation}
	\gamma \frac{d}{d t} \bm{\chi}_m(t) = - \xi_m \bm{\chi}_m(t) + \bm{\Gamma}_m(t) ,
	\label{eq::temporal_EOM}
\end{equation}	
are given by 
\begin{equation}
    \bm\chi_{m}(t) = \bm\chi_{m}(t_{0}) \: e^{-(t-t_{0})\xi_m/\gamma} + \frac{1}{\gamma} e^{-t\xi_m/\gamma} \int_{t_{0}}^{t} \bm\Gamma_{m}(t') \:  e^{t'\xi_m/\gamma} \: dt'  ,
\end{equation}
with the initial condition $\bm\chi_{m}(t_{0})$ at time $t_0$, and become  in the stationary state, $t_0 \rightarrow -\infty$, 
\begin{equation}
    \bm\chi_{m}(t) = \frac{1}{\gamma} e^{-t\xi_m/\gamma} \int_{-\infty}^{t} \bm\Gamma_{m}(t') \:  e^{t'\xi_m/\gamma} \: dt' .
\end{equation}
The correlation function of the latter is for $(m \ne 0)$ 
\begin{equation} 
	\lla \bm{\chi}_m(t) \cdot \bm{\chi}_n^{*}(t') \rra = \delta_{mn} \frac{3 k_BT}{\gamma} \tau_m e^{-|t-t'|/\tau_m} e^{i \omega_m|t-t'|} ,
	\label{eq::AMCF}
\end{equation}	
and for the translational mode, $m=0$,
\begin{equation}
	\braket{\bm{\chi}_0(t) \cdot \bm{\chi}_0(t_0)} = \braket{\bm{\chi}_0^2(t_0)} + \frac{6 k_BT }{\gamma } t .
\end{equation}

\section{Conformations of Passive Ring Polymers}

\begin{figure}[t]
	\centering
    \includegraphics[width=0.8\textwidth]{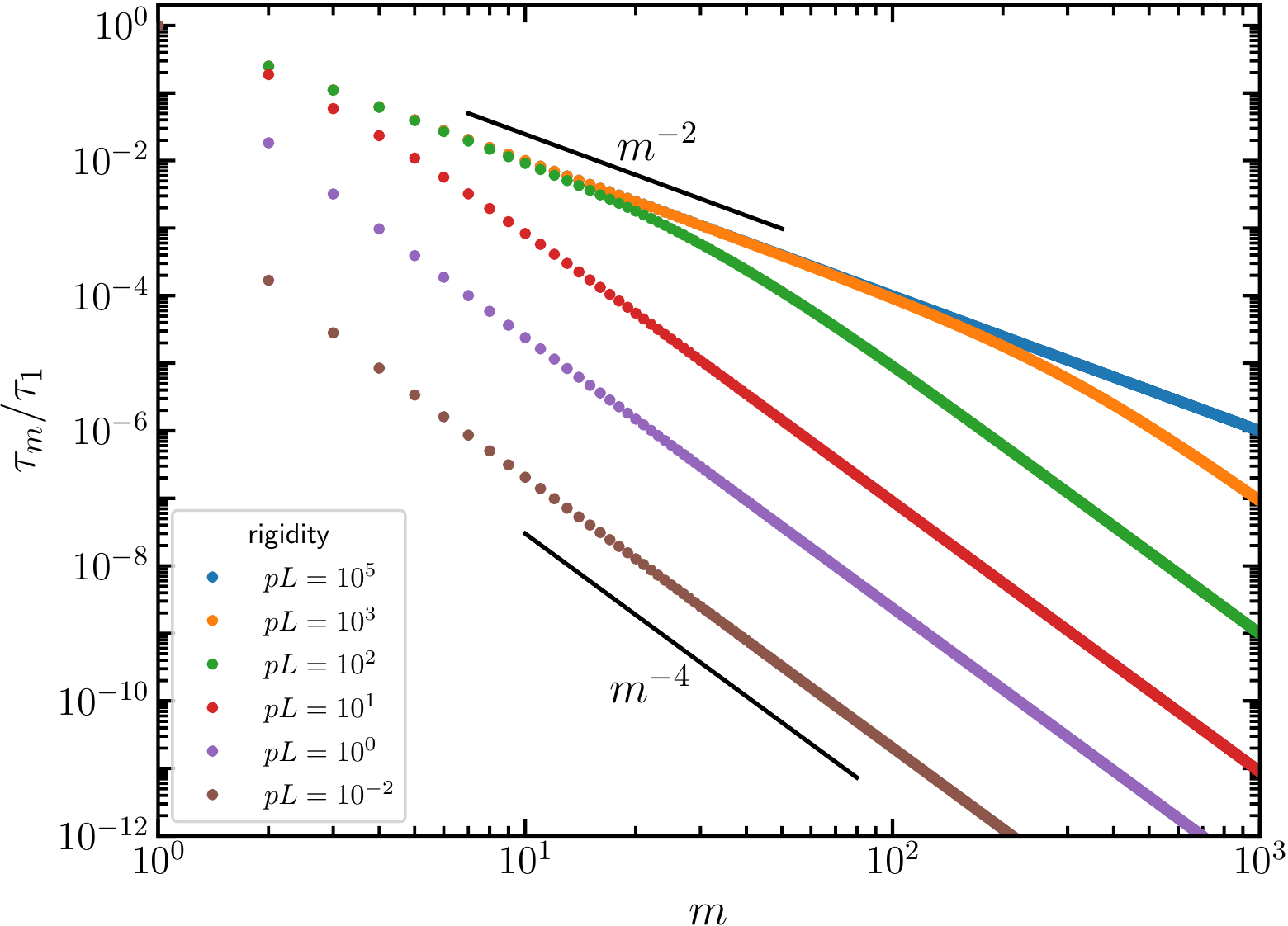}
	\caption{Normalized relaxation times $\tau_m / \tau_1$ as function of mode number $m$ for various $pL$ values, where  $\tau_1 = \gamma L^3 / (12\pi^2k_BTpL(\pi^2/(pL)^2+\mu))$ with $\mu = 2 \lambda/(3p)$. Flexible polymers, $pL \gg 1$, exhibit the mode-number dependence $1/m^2$, and semiflexible polymers, $pL \ll 1$, the dependence $1/m^4$. }
	\label{fig::RPM_con_pas_semi_tau}
\end{figure}	
\FloatBarrier

\subsection{Lagrangian Multiplier - Stretching Coefficient}
The ring-polymer inextensibility is captured via the local constraint of a unit mean-square tangent vector,
\begin{equation}
	\Bigg\langle \Bigg( \frac{\partial \bm{r}(s,t)}{\partial s} \Bigg)^2 \Bigg\rangle = 
	\sum_{\substack{m=-\infty \\m \ne 0}}^{\infty} \langle \bm{\chi}_m^2 \rangle \Big|\frac{\partial \phi_m(s)}{\partial s} \Big|^2
	= \frac{2}{pL} \sum_{m=1}^{\infty} \frac{1}{\pi^2 m^2/(pL)^2  + \mu}
	\equals^{!} 1.
\end{equation}	
Analytic evaluation of the infinite sum leads to implicit equations determining the stretching coefficient $\lambda = 3 p \mu/2$ at a given rigidity $pL$,
\begin{align}
	f(\mu;pL) &= \frac{1}{\sqrt{\mu}} \coth (pL\sqrt{\mu}) - \frac{1}{pL\mu} -1 \equals^{!}0 ,  \hspace{1.45cm} pL > 3,
	\\
	f(\mu;pL) &= \frac{1}{pL|\mu|} - \frac{1}{\sqrt{|\mu|}} \cot \small(pL\sqrt{|\mu|}\small) -1 \equals^{!}0 ,
	\hspace{0.95cm} pL < 3.
\end{align}	
In the flexible case, $pL \gg 1$, $\mu$ is unity, and in the semiflexible regime (SFR), $pL \ll 1$, $\mu$ can be approximated by
\begin{equation}
	\mu \approx - \frac{\pi^2}{(pL)^2} \Big( 1 - \frac{2pL}{\pi^2} \Big).
	\label{eq::mu_semi}
\end{equation}	
Figure~\ref{fig::RPM_con_pas_semi_LM_MSRD} displays the dependence of $\mu$ on $pL$.

\subsection{Mean-Square Ring Diameter}
The mean-square ring diameter $\langle \bm{r}_d^2 \rangle = \langle (\bm{r}(L/2)-\bm{r}(0))^2 \rangle$  is given by
\begin{align} \label{eq::msq_end}
		\frac{\langle \bm{r}_d^2 \rangle}{L^2} 
		=\frac{8}{L^3} \sum_{m=1}^{\infty} \langle \bm{\chi}_{2m-1}^2 \rangle
		 = \frac{2}{\pi^2pL} \sum_{m=1}^{\infty} \frac{1}{\pi^2 (2m-1)^4/(pL)^2 + \mu (2m-1)^2}
		=\frac{1}{4pL\mu} \Big[1- \frac{2}{pL\sqrt{\mu}} \tanh(pL\sqrt{\mu}/2) \Big].
\end{align}	
The dependence on $pL$ is presented in Fig.~\ref{fig::RPM_con_pas_semi_LM_MSRD}.

\begin{figure}[t]
	\centering
    \includegraphics[width=0.8\textwidth]{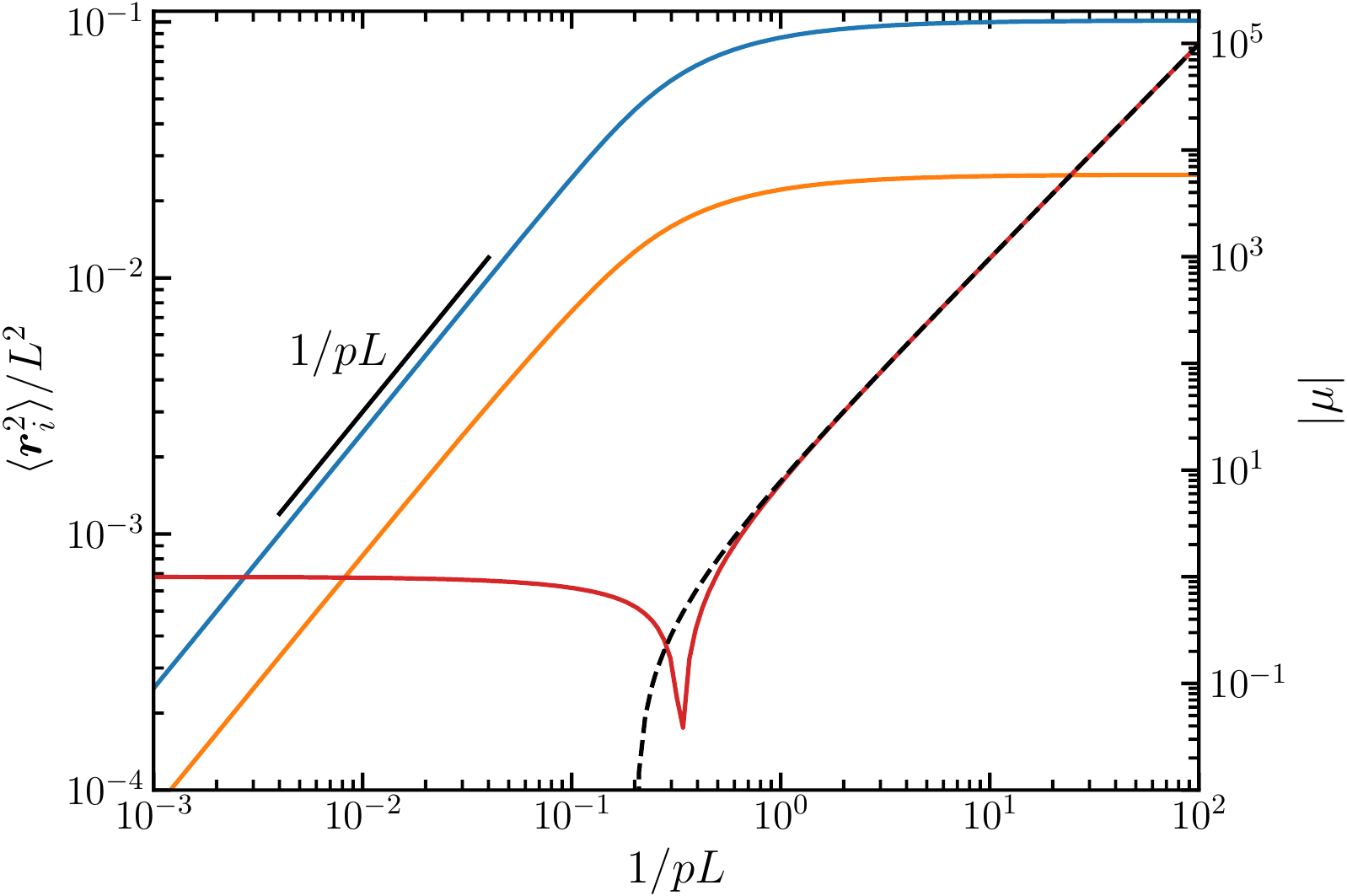}
	\caption{Normalized mean-square ring diameter $\langle \bm{r}_d^2 \rangle$, blue line, normalized mean-square radius of gyration $\langle \bm{r}_g^2 \rangle$, yellow line, (left-hand axis) and the magnitude of the stretching coefficient $\mu$, red line, (right-hand axis) as function of the inverse stiffness $1/pL$. The black line shows a power law with the indicated inverse rigidity dependence. The dashed back line corresponds to the approximation of the stretching coefficient in the semiflexible regime, $pL \ll 1$, Eq.~\eqref{eq::mu_semi}.}
	\label{fig::RPM_con_pas_semi_LM_MSRD}
\end{figure}	
\FloatBarrier

\subsection{Mean-Square Radius of Gyration}

The radius of gyration is given by (see Fig.~\ref{fig::RPM_con_pas_semi_LM_MSRD}) 
\begin{align} \label{eq::gyrat}
    \begin{split}
    	\frac{\braket{\bm{r}_g^2}}{L^2} &= \frac{1}{2L^4} \int_{0}^{L} \int_{0}^{L} \left\langle (\bm{r}(s_1) - \bm{r}(s_2))^2 \right\rangle \: ds_1 ds_2 = \frac{2}{L^3} \sum_{m=1}^{\infty} \braket{\bm{\chi}^2_m}
    	\\
    	& = \frac{1}{2\pi^2pL} \sum_{m=1}^{\infty} \frac{1}{ \pi^2 m^4/ (pL)^2 + \mu m^2}
    	= \frac{1}{12 pL \mu} \Big[1 + \frac{3}{(pL)^2\mu} - \frac{3}{pL\sqrt{\mu}} \coth(pL \sqrt{\mu}) \Big] .
	\end{split}
\end{align}

\subsection{Ring Polymer Size Limits}

In the flexible limit, $pL \rightarrow \infty$ and $\mu = 1$, Eqs.~\eqref{eq::msq_end} and \eqref{eq::gyrat} become
\begin{equation}
    \frac{\langle \bm{r}_d^2 \rangle}{L^2} \approx \frac{1}{4pL} , 
    \hspace{2cm}  
    \frac{\braket{\bm{r}_g^2}}{L^2} \approx \frac{1}{12pL} ,
\end{equation}
and in the semiflexible limit, $pL \rightarrow 0$, 
\begin{equation}		
	\frac{\langle \bm{r}_d^2 \rangle}{L^2} \approx \frac{1}{\pi^2} ,
	\hspace{2cm}
	\frac{\braket{\bm{r}_g^2}}{L^2} \approx \frac{1}{4\pi^2} .
\end{equation}	
The presented conformational analysis of the passive ring polymer is a recapitulation of Ref.~[S3].

\subsection{Spatial Tangent Vector Correlation Function}
The spatial correlation function of the tangent vector, $\bm{u}(s,t) = \partial \bm{r}(s,t)/\partial s$,  defined as
\begin{align} \label{eq::corr_tangent}
	\langle \bm{u}(s,t) \cdot \bm{u}(s',t)\rangle = \frac{2}{L} \sum_{m=1}^{\infty} \langle \bm{\chi}_m(t) \cdot \bm{\chi}_m^{*}(t) \rangle \: k_m^2 \: Re \Big[e^{ik_m(s-s')} \Big] = \frac{2}{pL} \sum_{m=1}^{\infty} \frac{\cos(2\pi m (s-s') /L)}{\pi^2m^2/(pL)^2 + \mu} ,
\end{align}	
is zero for flexible polymers ($s\neq0$)  and approaches a cosine function for very stiff rings, as displayed in Fig. \ref{fig::RPM_con_pas_semi_LM_STVC}.
\begin{figure}[t]
	\centering
    \includegraphics[width=0.8\textwidth]{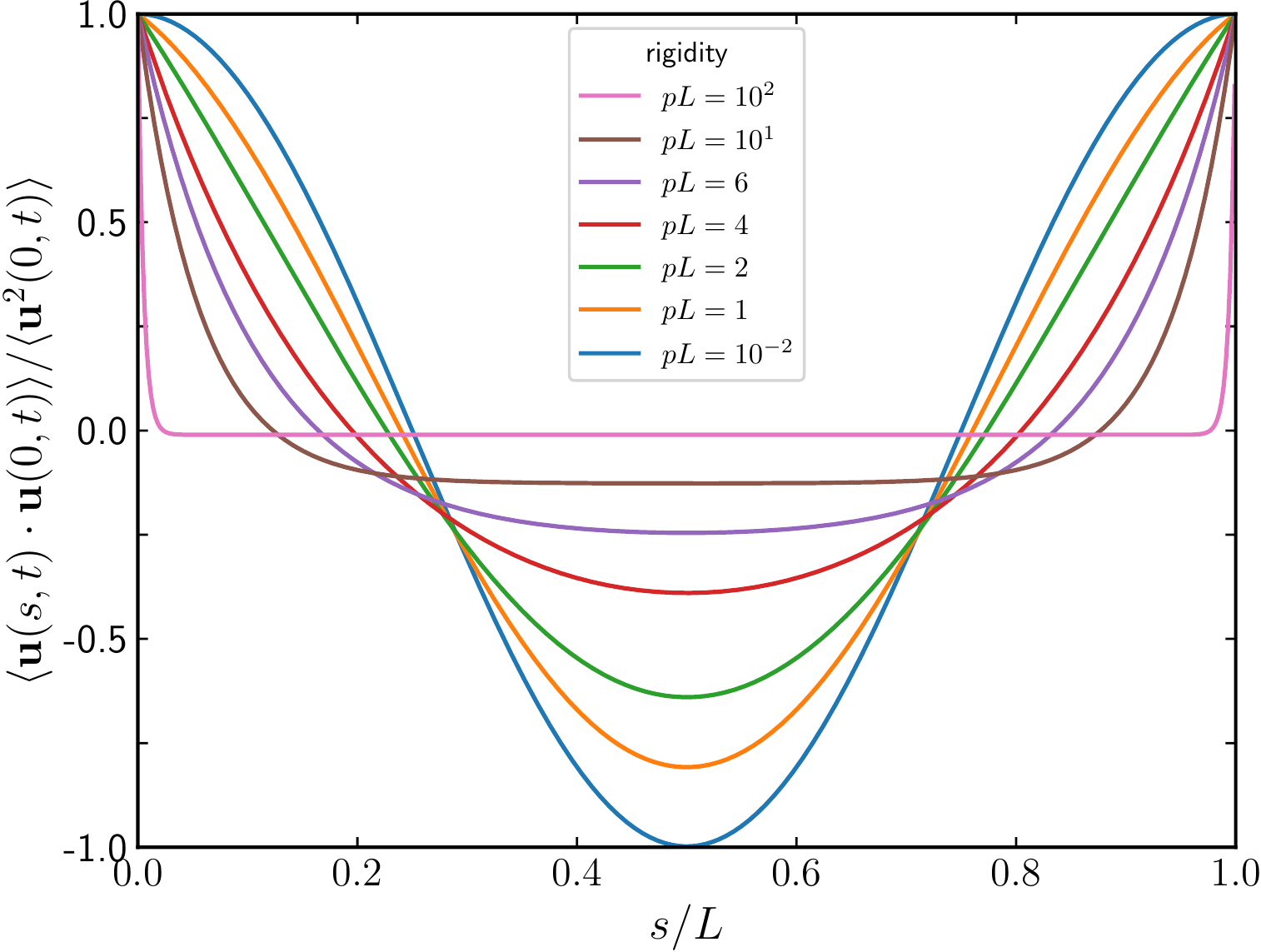}
	\caption{Normalized spatial tangent vector correlation function, Eq.~\eqref{eq::corr_tangent},  as a function of position $s / L$ for different $pL$ values.}
	\label{fig::RPM_con_pas_semi_LM_STVC}
\end{figure}	
\FloatBarrier

\section{Dynamics of Active Polar Ring Polymers}
\subsection{Mean-Square Displacement}
The dynamics of the ring's center-of-mass $\bm{r}_{cm}(t)=\int_{0}^{L}\bm{r}(s,t)ds/L=\bm{\chi}_0(t)\phi_0$ is characterized by its mean-square displacement (MSD), which yields
\begin{equation}
	\langle \Delta\bm{r}_{cm}^2(t) \rangle = \langle (\bm{r}_{cm}(t)-\bm{r}_{cm}(0))^2 \rangle = 6 D_0 t,
\end{equation}	
with $D_0 = k_BT/(\gamma L)$ the diffusion coefficient. By integration of the equation of motion over the polymer contour, all internal forces vanish, and the center-of-mass mean-square displacement is independent of activity.  

The MSD of a point $\bm r(s,t)$ in the center-of-mass reference frame,  $\langle \Delta \bm{r}^2(t) \rangle =  \langle (\bm{r}(s,t)-\bm{r}(s,0))^2 \rangle - 6 D_0 t$, is given by
\begin{align} \label{eq::msd_cm_general}
	\frac{\langle\Delta \bm{r}^2(t) \rangle}{L^2} &= \frac{12k_BT}{\gamma L^3} \sum_{m=1}^{\infty} \tau_m \left(1 - \cos \left(\omega_m t \right) \:  e^{- t\tau_m}\right) .
\end{align}	
In the limit of a  \textbf{flexible} ring, $pL \gg 1$, Eq.~\eqref{eq::msd_cm_general} reduces to
\begin{equation}
	\frac{\langle\Delta \bm{r}^2(t)\rangle}{L^2} \approx \frac{1}{\pi^2pL} \sum_{m=1}^{\infty} \frac{1}{m^2} \Big(1- \cos \Big(\frac{m}{6 \pi} \frac{Pe}{pL} \frac{t}{\tau_1} \Big) \: e^{-m^2t/\tau_1} \Big) .
\end{equation}	
For $t / \tau_1 \ll 1$ all modes contribute to the sum-over-modes, and the sum can be replaced by an integral. The substitution $z^2 = m^2 t / \tau_1$ yields then 
\begin{equation}
	\frac{\langle \Delta \bm{r}^2(t)\rangle}{L^2} \approx \frac{1}{\pi^2pL} \Bigg[ \sqrt{\frac{t}{\tau_1}} \underbrace{\int_{0}^{\infty} \frac{1 - e^{-z^2}}{z^2} dz}_{=\sqrt{\pi}} \: + \: \frac{1}{2(6\pi)^2} \frac{Pe^2}{(pL)^2} \Big(\frac{t}{\tau_1}\Big)^{3/2} \underbrace{\int_{0}^{\infty} e^{-z^2} dz}_{=\sqrt{\pi}/2} \Bigg] 
\end{equation}	
for $t/\tau_1 \ll 1/(\omega_1 \tau_1)^2$, with the well-known subdiffusive time dependence $t^{1/2}$, and a $t^{3/2}$ crossover to an activity enhanced diffusive regime. The latter appears on time scales $1/(\omega_1 \tau_1)^2 < t/\tau_1 \ll 1/(\omega \tau_1)$, where $t/\tau_1 \ll 1$, and the substitution $z = m Pe \: t / (6 \pi pL \tau_1)$ yields
\begin{equation}
	\frac{\langle \Delta \bm{r}^2(t) \rangle}{L^2} \approx \frac{1}{\pi^2pL} \sum_{m=1}^{\infty} \frac{1}{m^2} \Big[1 - \cos \Big(\frac{m}{6 \pi} \frac{Pe}{pL} \frac{t}{\tau_1} \Big) \Big] \approx \frac{1}{6\pi^3} \frac{Pe}{(pL)^2} \frac{t}{\tau_1} \underbrace{\int_{0}^{\infty} \frac{1-\cos(x)}{x^2} \: dz}_{=\pi/2} .
\end{equation}	
In the limit of a \textbf{semiflexible} ring, $pL \ll 1$,  Eq.~\eqref{eq::msd_cm_general} becomes
\begin{equation} \label{eq::msd_semi_full}
	\frac{\langle\Delta \bm{r}^2(t) \rangle}{L^2} \approx  \frac{1}{2 \pi^2} \left(1-\cos(\omega_1 t)e^{-t/\tau_1}\right) +
	\frac{pL}{\pi^4} \sum_{m=2}^{\infty} \frac{1 - \cos \left(m Pe /(12\pi) \, t/\tau_1 \right) \: e^{- \pi^2 m^4/(2 pL) \,  t/\tau_1}}{m^4} ,
\end{equation}	
with the longest relaxation time $\tau_1 = \gamma L^3/(24 \pi^2 k_BT)$, which is independent of $pL$ [S1].  A similar value of the rotational relaxation time, up to a factor of $2$, has been derived by Bixon et al. via a linear response theory approach [S4].  The first term on the right-hand side of Eq.~\eqref{eq::msd_semi_full} describes the rotational motion [S5].
At short times  $t / \tau_1 \ll \mathrm{min}\{pL, (pL)^{1/5}/(\omega_1 \tau_1)^{8/5}\}$, the substitution  $z^2 = \pi^2 m^4 t/(2pL\tau_1)$ and integration yields
\begin{equation}
	\frac{\langle \Delta \bm{r}^2(t) \rangle}{L^2} \approx \frac{(pL)^{1/4}}{2^{7/4}\pi^{5/2}} \Big(\frac{t}{\tau_1}\Big)^{3/4} \underbrace{\int_0^{\infty} \frac{1-e^{-z^2}}{z^{5/2}} dz}_{=2.42} \hspace{0.2cm} 
\end{equation}
as for a passive ring, with the well-known $t^{3/4}$ dependence {S1,S6}.
For times $t/\tau_1 > \mathrm{min}\{pL, (pL)^{1/5}/(\omega_1 \tau_1)^{8/5}\}$, the internal ring dynamics is dominated by the first mode (first term on the right-hand side of Eq.~\eqref{eq::msd_semi_full}), and Taylor expansion gives 
\begin{align}
\frac{\langle\Delta \bm{r}^2(t) \rangle}{L^2} \approx \frac{Pe^2}{(24 \pi^2)^2} \Big( \frac{t}{\tau_1} \Big)^2 ,
	\label{align::RPM_con_act_semi_MSD_cmf_ITA}
\end{align}	
for the range $(pL)^{1/5}/(\omega_1 \tau_1)^{8/5} < t/\tau_1 \ll 1/(\omega_1 \tau_1)$. The long-time limit of the internal dynamics converges to $\langle \Delta \bm{r}^2(t) \rangle \xrightarrow[]{t \rightarrow \infty} 2 \langle \bm{r}_g^2 \rangle$ for any P\'eclet number and stiffness  $pL$. \\

The contribution of the mode $m=1$ to the MSD in the center-of-mass reference frame for different rigidities $pL$ is displayed in Fig.~\ref{fig::RPM_con_act_MSD_mode_comparison}. For flexible rings, all modes contribute and the neglect of the first mode $m=1$ results in an average deviation of $60\%$ from the exact result in the oscillatory regime and the long time limit, yet still showing oscillations as expected for reptation motion. However, the rotational motion of semiflexible rings is solely determined by the longest relaxation time, the rotation relaxation time $\tau_1$ (see Fig. \ref{fig::RPM_con_act_MSD_mode_comparison}$(b)$).

\begin{figure}[t]
	\centering
    \includegraphics[width=1.0\linewidth]{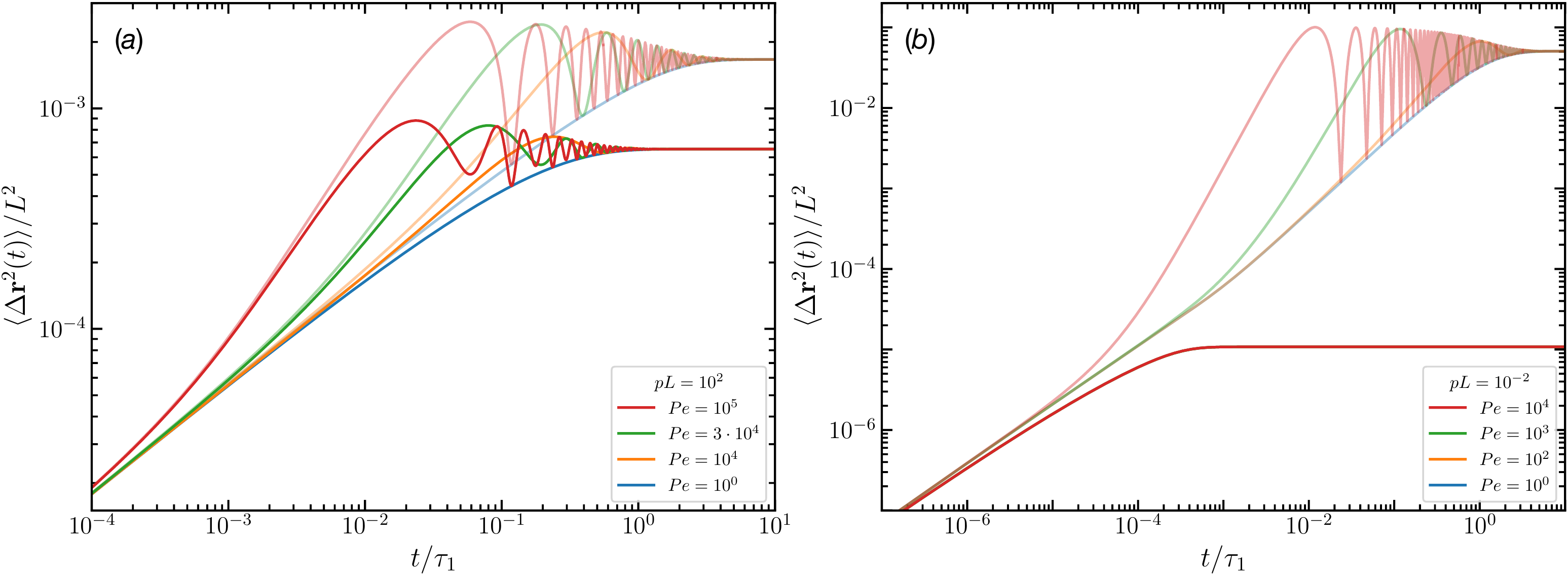}
	\caption{Mean-square displacement in the center-of-mass reference frame, $\langle \Delta \bm{r}^2(t)\rangle$, as function of time $t/\tau_1$ for (a) flexible, $pL = 10^{2}$ and (b) stiff, $pL = 10^{-2}$, rings for various P\'eclet numbers $Pe$, as in Fig. 2 of the main text. The transparent curves show the full MSD determined by all modes, while the colour-saturated ones display the MSD without the first mode contribution $m=1$.}
	\label{fig::RPM_con_act_MSD_mode_comparison}
\end{figure}
\FloatBarrier 

The transition from the activity-enhanced diffusion of flexible to the ballistic regime of semiflexible rings with increasing stiffness is shown in Figure~\ref{fig::RPM_con_act_semi_MSD_cmf_adj_pL_comparison_1}. Curves of semiflexible rings with $pL < 10^{-2}$ are on time scales $t/\tau_1 > (pL)^{1/5}/(\omega_1 \tau_1)^{8/5}$, indistinguishable from that for $pL = 10^{-2}$, since $\tau_1$ is independent of $pL$,  only time regime extents to shorter times.

\begin{figure}[b]
	\centering
    \includegraphics[width=0.75\textwidth]{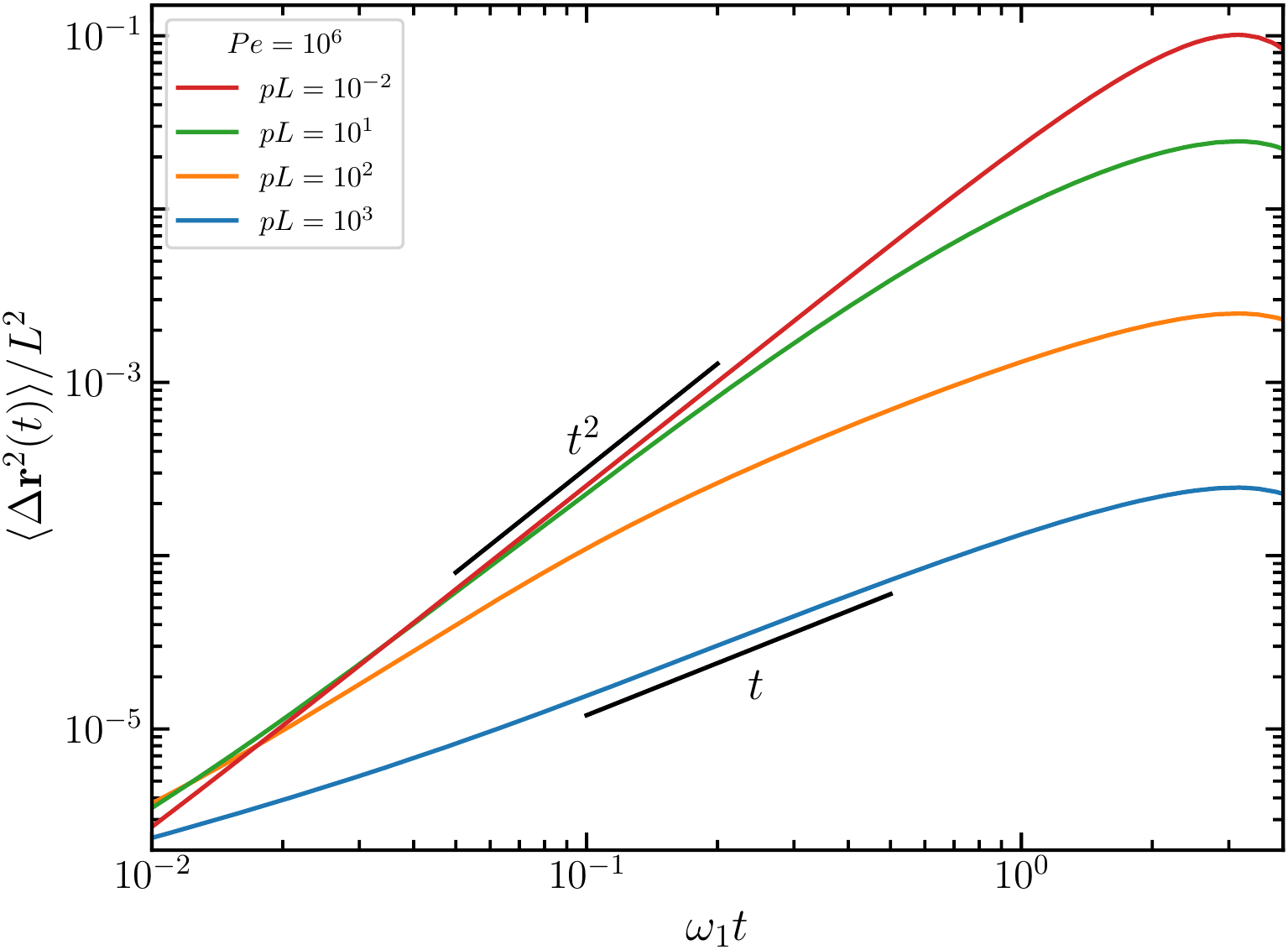}
	\caption{Normalized mean-square displacement in the center-of-mass frame, $\langle \Delta \bm{r}^2(t) \rangle / L^2 $, as function of time $\omega_1 t$ for P\'eclet number $Pe = 10^{6}$ and different rigidities $pL$. The black lines show power laws with the indicated time dependence.}
	\label{fig::RPM_con_act_semi_MSD_cmf_adj_pL_comparison_1}
\end{figure}	
\FloatBarrier

\subsection{Temporal Autocorrelation Function of the Ring Diameter}

The temporal correlation of the ring diameter, $\langle \bm{r}_d(t) \cdot \bm{r}_d(0) \rangle$, is given by
\begin{equation}
	\frac{\langle \bm{r}_d(t) \cdot \bm{r}_{d}(0) \rangle}{L^2} = \frac{8}{L^3} 	\sum_{m=1}^{\infty} Re \Big[ \langle \bm{\chi}_{2m-1}(t) \cdot \bm{\chi}_{2m-1}^{*}(0)  \rangle \Big] = \frac{24k_BT}{\gamma L^3} \sum_{m=1}^{\infty} \tau_{2m-1} \: \cos( \omega_{2m-1} t) \: e^{-t / \tau_{2m-1}}.
	\label{eq::OVC}
\end{equation}	
For flexible rings, the correlation becomes 
\begin{equation}
	\frac{\langle \bm{r}_d(t) \cdot \bm{r}_{d}(0) \rangle}{L^2} \approx \frac{2}{\pi^2pL} \cos\left(\frac{1}{6\pi} \frac{Pe}{pL} \frac{t}{\tau_1}\right)e^{-t/\tau_1} \ ,
	\hspace{1.55cm} 
	\tau_1 \approx \frac{\gamma L^3}{12\pi^2k_BTpL} ,
\end{equation}	
and depends on rigidity $pL$. In case of semiflexible rings, the correlation function, using Eq.~\eqref{eq::mu_semi}, is 
\begin{equation}
    \hspace{-0.4cm}
	\frac{\langle \bm{r}_d(t) \cdot \bm{r}_{d}(0) \rangle}{L^2} \approx \frac{1}{\pi^2} \cos\left(\frac{Pe}{12\pi} \frac{t}{\tau_1}\right)e^{-t/\tau_1} \ , 
	\hspace{2.3cm} 
	\tau_1 \approx \frac{\gamma L^3}{24\pi^2k_BT} ,
\end{equation}	
which is independent of $pL$. In any case,  $\langle \bm{r}_d(t) \cdot \bm{r}_{d}(0) \rangle$ depends on $Pe$ for $t>0$. The largest deviation between the correlation function including all modes and the mode $m=1$ only appears for $t=0$, i.e., for the equilibrium value. This deviation is presented in Figure~\ref{fig::RPM_con_act_semi_OVC_adj_pL_t_zero}, where $\langle \Delta \bm{r}_d^2(0) \rangle = \langle \bm{r}_d^2(0) \rangle - \langle \bm{r}_d^2(0) \rangle_{m=1}$ is the difference between  the equilibrium mean-square ring diameter including all modes, $\langle \bm{r}_d^2(0) \rangle$, and that with the first mode only, $\langle \bm{r}_d^2(0) \rangle_{m=1}$. The largest deviation of around $20\%$ is obtained for flexible rings and the deviation decreases with increasing stiffness. 

\begin{figure}[t!]
	\centering
    \includegraphics[width=0.8\textwidth]{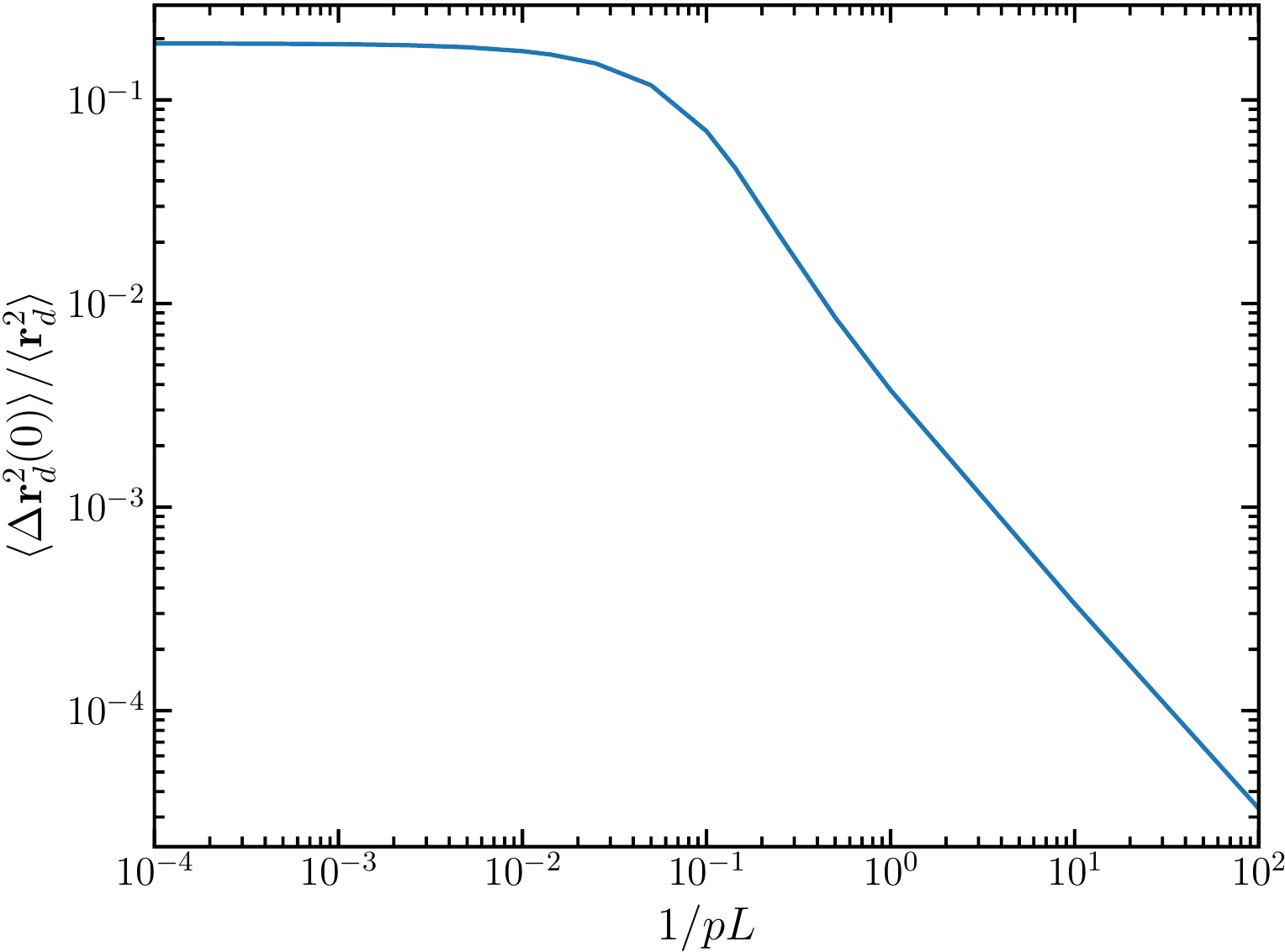}
	\caption{Normalized difference between  the equilibrium mean-square ring diameter $\langle \Delta \bm{r}_d^2(0) \rangle = \langle \bm{r}_d^2(0) \rangle - \langle \bm{r}_d^2(0) \rangle_{m=1}$ including all modes, $\langle \bm{r}_d^2(0) \rangle$, and that with the first mode only, $\langle \bm{r}_d^2(0) \rangle_{m=1}$, as a function of  $1/(pL)$.}
	\label{fig::RPM_con_act_semi_OVC_adj_pL_t_zero}
\end{figure}	

\newpage

\section{Brownian Dynamics Simulation}

In simulations, a ring polymer composed of $N$ monomers is considered, which obey the overdamped equations of motion 
\begin{align}  \label{eq:langevin_disc}
\hat \gamma \dot {\bm r}_i (t) = \bm F_i^a(t) + \bm F_i(t) +  \bm \Gamma_i(t)  ,
\end{align} 
with the friction coefficient $\hat \gamma$, the active force $\bm F_i^a$, intramolecular forces $\bm F_i$, and stochastic force $\bm \Gamma_i$. The tangential active force $\bm F_i^a$ is given by 
\begin{align}
\bm F_i^a = \frac{F_a}{2l}  \left( \bm R_{i+1}   +  \bm R_{i}  \right)  = \frac{F_a}{2l}  \left(  \bm r_{i+1} - \bm r_{i-1}  \right) ,
\end{align}
where $\bm R_{i+1} = \bm r_{i+1} - \bm r_i$ is the bond vector, $l$ is the bond length, and $F_a$ the constant magnitude  of the active force  For semiflexible phantom polymers, the intramolecular forces  follow from the bond, $U_l$, and bending, $U_b$, potentials  
\begin{align} \label{eq:pot_bond}
U_l = & \ \frac{\kappa_l}{2} \sum_{i=1}^{N} \left(| \bm R_{i+1}| -l \right)^2  ,\\ \label{eq:pot_bend}
U_b = & \ \frac{\kappa_b}{2} \sum_{i=1}^{N} \left(\bm R_{i+1} - \bm R_i \right)^2 .
\end{align}
$\kappa_l$ and $\kappa_b$ are the strengths of the potentials, 
and $\bm r_{N+1} \equiv \bm r_1, \ \bm r_{-1} \equiv \bm r_N$. $\bm \Gamma_i$ is  Gaussian and Markovian stochastic processes with zero mean and the second moments
\begin{align}
\lla \bm \Gamma_{i}(t) \cdot \bm \Gamma_j(t')  \rra = 6 k_BT \hat \gamma \delta_{ij} \delta (t-t') . 
\end{align}
Activity is characterized by the P\'eclet number $Pe=F_a l/(k_BT)$.    

The equations of motion are integrated via the Euler method with a time step of $\Delta t = 2 \times 10^{-7}  l^2\gamma/(k_BT)$, and $\kappa_l$ is set to $\kappa_l= 10^4 k_BT/l^3$, which insures that $|\bm R_i| = l$. 

Additionally, we performed simulations taking the inertia term, $m \ddot{\bm r}_i (t)$, into account. Noteworthy, such rings exhibit a pronounced swelling, rather than a collapse. Hence, inertia leads to a rather different conformations and dynamics.

\section{Movies} 

\noindent {\bf Supplemental Movie M-1} \\
Reptation motion of a flexible ring with $N=50$ monomers for the P\'eclet number  $Pe=1$. Half of the monomers are colored in red and blue, respectively. \\[1ex]
\noindent {\bf Supplemental Movie M-2} \\
Reptation motion of a semiflexible ring with $N=50$ monomers for the P\'eclet number  $Pe=1$. The stiffness parameter in Eq.~\eqref{eq:pot_bend} is chosen as $\kappa_b= 10 k_BT/l^3$.  Half of the monomers are colored in red and blue, respectively. \\[1ex]
{\bf Supplemental Movie M-3} \\
Tank-treading motion  of a stiff ring with $N=50$ monomers for the P\'eclet number  $Pe=1$. The stiffness parameter in Eq.~\eqref{eq:pot_bend} is chosen as $\kappa_b= 10^5 k_BT/l^3$. Half of the monomers are colored in red and blue, respectively. \\[1ex]

\noindent [S1] L. Harnau, R. G. Winkler, and P. Reineker,  J. Chem. Phys. {\bf 102}, 7750 (1995). \\
\noindent  [S2] R. G. Winkler, J. Chem. Phys. {\bf 118}, 2919 (2003). \\
\noindent  [S3] S. M. Mousavi, G. Gompper, and R. G. Winkler, J. Chem. Phys. {\bf 150}, 064913 (2019). \\
\noindent [S4] 	M. Bixon and R. Zwanzig, J. Chem. Phys. {\bf 68}, 1896 (1978). \\
\noindent [S5]	R. G. Winkler, J. Chem. Phys. {\bf 127}, 054904 (2007).\\
\noindent [S6] E. Farge and A. C. Maggs, Macromolecules {\bf 26}, 5041 (1993).
\end{widetext}

\end{document}